\documentclass[]{tJOT2e}
\begin{document}
\doi{10.1080/14685248.YYYYxxxxxx}
 \issn{1468-5248}
 \jvol{00} \jnum{00} \jyear{2011}

\markboth{S. Zammert and B. Eckhardt}{Journal of Turbulence}

\articletype{Paper}

\title{Harbingers and latecomers - The order of appearance of exact coherent structures in plane Poiseuille flow}

\author{Stefan Zammert$^{\rm a}$$^{\ast}$\thanks{$^\ast$Corresponding author. Email: Stefan.Zammert@physik.uni-marburg.de
\vspace{6pt}} and Bruno Eckhardt$^{\rm b}$\\\vspace{6pt}  $^{\rm a}${\em{Fachbereich Physik, Philipps-Universit\"{a}t Marburg, \\ Renthof 6, D-35032 Marburg, Germany}};\\
$^{\rm b}${\em{J.M. Burgerscentrum, Delft University of Technology, \\ 2628 CD Delft, The Netherlands}}\\\vspace{6pt}}

\maketitle

\begin{abstract}
The transition to turbulence in plane Poiseuille flow (PPF) is connected with the presence of  
exact coherent structures. In contrast to other shear flows, PPF has a number
of different coherent states that are relevant for the transition. We here discuss the different 
states, compare the critical Reynolds numbers and optimal wavelengths for their appearance, 
and explore the differences between flows operating at constant mass flux or at constant pressure
drop. The Reynolds numbers quoted here are based on the mean flow velocity and 
refer to constant mass flux, the ones for constant pressure drop are always higher. 
The Tollmien-Schlichting waves bifurcate subcritically from the laminar profile at $Re=5772$
and reach down to $Re=2609$ (at a different optimal wave length). Their localized counter part
bifurcates at the even lower value $Re=2334$.
Three dimensional exact solutions appear at much lower Reynolds numbers. We describe
one exact solutions that is spanwise localized and has a critical Reynolds number of  $316$. 
Comparison to plane Couette flow suggests that this is likely the lowest Reynolds number 
for exact coherent structures in PPF.
Streamwise localized versions of this state require higher Reynolds numbers, with the lowest 
bifurcation occurring near $Re=1018$. 
\end{abstract}


\section{Introduction}
\label{main}
Within the last decade many fully 3d exact coherent states of the Navier-Stokes equations have been identified.
They are important for understanding the transition to turbulence in various shear flows, for determining the thresholds
for the transition, and for characterizing the subsequent evolution towards fully developed turbulence \cite{Eckhardt2007,Eckhardt2008b}.
Coherent structures provide a scaffold for the turbulent time evolution because they are embedded in the turbulent dynamics \cite{Kawahara2001} and form a network of heteroclinic connections  \cite{Halcrow2009}. Tracking exact coherent states also provided insights
into the formation of the chaotic saddles and transient turbulence in  plane Couette, pipe, and plane Poiseuille flow (PPF) \cite{Kreilos2012,Avila2013,Zammert2015}.
Special coherent structures, so called edge states \cite{Skufca2006}, are important for the transition
to turbulence since their stable manifold separates the state space in a part with turbulence and another 
one where initial conditions relaminarize directly.

Among the various shear flows that have been studied, PPF is special because it also has a 
linear instability of the laminar profile which triggers Tollmien-Schlichting (TS) waves \cite{Henningson}. 
The extend to which they contribute to the
observed transition then depends on the order of appearance of the different states. As we will see,
some states are \textit{harbingers} that appear well below the 
experimentally observed transition, and others are \textit{latecommers}, appearing well above the transition.
Harbingers prepare the experimentally observed transition to turbulent dynamics, and latecomers add additional
states and degrees of freedom at higher Reynolds numbers. 

While critical Reynolds numbers for linear instabilities of the laminar profile are well defined and unique, this is not
the case for the subcritical transitions to be discussed here. For the subcritical case, it does make a difference whether the
system is run under conditions of constant mass flux or prescribed pressure or perhaps constant energy input
 \cite{Yosuke2013}. All exact coherent states have a fixed relation between pressure drop and mean flow rate, 
but the quest for the bifurcation point focusses on different projections and hence gives different critical Reynolds numbers
depending on whether the flow rate or the pressure drop or some other quantity are kept constant.

The setting of operating conditions also affects the choice of Reynolds numbers. To be specific, we define 
\begin{equation}
 Re_{B}=\frac{3 U_{B}d }{ 2\nu},
 \label{Re_B}
\end{equation}
for the case of constant mean bulk velocity $U_{B}$. The other parameters are $d$, half the channel width and $\nu$, the kinematic viscosity.
If the pressure gradient is fixed, we define the pressure based Reynolds number, 
\begin{equation}
 Re_{P}=\frac{U_{cl,P}d}{\nu},
 \label{Re_P}
\end{equation}
where $U_{cl,P}=d^2 (dP/dx) / 2\nu$ is the laminar centerline velocity for this value of the pressure gradient.
The numerical factors in  (\ref{Re_B})  are chosen such that the two Reynolds numbers agree for the case of 
a laminar profile, $Re_B=Re_P$. 
For 3d coherent states they are usually different. 
Therefore, even if the laminar state coincides, the state space of the system at different operating conditions may be 
different, because in one case some exact coherent structure might already exist that are not yet present in the other case. 

For direct numerical simulation we used \textit{Channelflow}-code (www.channelflow.org) \cite{J.F.Gibson2012}.
Exact coherent structures were identified and continued in Reynolds number using the Newton hookstep 
 \cite[see e.g.][]{Viswanath2007}  continuation methods \cite[see e.g.][]{Dijkstra2013}
included in the \textit{Channelflow}-package. 

\section{Spatially extended and localized Tollmien-Schlichting waves}
The laminar state of PPF has a subcritical instability. 
The exact values for the critical Reynolds number and wavelength have
remained unclear
 \cite{Heisenberg1924,Lin1945,Thomas1953} until the issue was finally settled
by Orszag \cite{Orszag1971}, who calculated a critical Reynolds number of $5772.22$
for a critical wavelength of $1.96\pi$ using spectral methods. Chen \& Joseph \cite{Chen1973} then demonstrated the existence 
of two-dimensional travelling wave solutions, so-called finite amplitude TS-waves,
bifurcating subcritically from the laminar flow. These two-dimensional exact coherent structures were later 
studied by Zahn et al. \cite{Zahn1974}, Jimenez \cite{Jimenez1990a} and Soibelman \& Meiron \cite{Soibelman1991}. 
Estimates for the lowest Reynolds number of the turning points were first given by Grohne \cite{Grohne1969} and Zahn et al. \cite{Zahn1974}. 
Using a truncation after one or two modes they found that the traveling waves appear slightly above $Re=2700$.
Their result were qualitatively confirmed in later studies using a higher number of modes \cite{Herbert76}.  

In our numerical simulations we use the \textit{channelflow}-code with resolution $N_{x} \times N_{y} \times N_{z}=80 \times 97 \times 4$. 
The streamwise and spanwise resolution is chosen sufficiently fine to ensure that a further increase
does not change the obtained Reynolds numbers. The code is set up to solve the fully 3d problem, 
however, it can be reduced to an effectively 2d code with the
choice of $N_z=4$, since all the modes with $k_z\ne0$ enter with zero amplitude. We therefore can stay within 
the same algorithmic framework and code and can exploit all the modules of \textit{channelflow}.

\begin{figure}[t]\vspace*{4pt}
\centering
\includegraphics[]{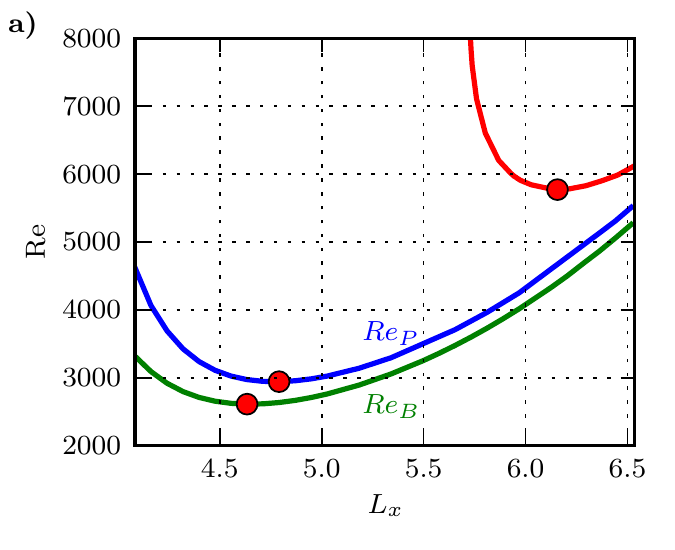}\includegraphics[]{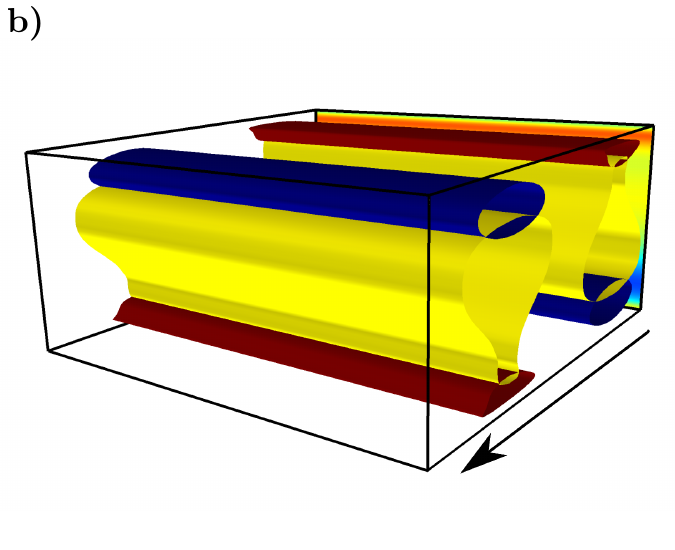}
\caption{Tollmien-Schlichting waves in PPF.
(a) Instability of the laminar profile (red) and existence regions for spatially extended TS waves.
The green and blue lines show the Reynolds number of the turning point vs. streamwise wavelength $L_{x}$
for the constant mean flow (green) and constant pressure gradient (blue), respectively. 
The minima of the curves are marked by a red circle: $Re=5772$ and $\lambda=6.16$ for the linear instability of the
laminar profile, $Re_B= 2609$ and $\lambda=4.65$ for constant mass flux and $Re_P=2941$ and $\lambda=4.81$
for constant pressure drop.
(b) Visualization of the TS wave for the minimal Reynolds number at constant mass flux. 
The direction of the flow is indicated by the black arrow. 
An iso-surface $0.01$ of the Q-vortex criterion is shown in yellow. Iso-surfaces $\pm0.12$ of the streamwise velocity (deviation from laminar)
are shown ind blue and red. 
}\vspace*{-6pt}
\label{fig_BifTSwaves}
\end{figure}

We identify the traveling waves that bifurcate from the laminar flow (referred to as $TW_{TS}$ in the following) using a 
Newton-method and continue them in Reynolds number with the continuation methods within \textit{channelflow}.
Figure \ref{fig_BifTSwaves}a) shows the bifurcation diagram versus streamwise wavelength.
The stability curve for the laminar profile is the same 
for the constant pressure gradient and constant mass flux case, and has a minimum at $Re=5772$
and a wavelength of $6.16$.
The bifurcation is subcritical and reaches to lower Reynolds numbers, but these points then depend on the 
operating conditions: the minimal values are $Re_{B} = 2609$ and $Re_{P} = 2941$.
The critical wavelengths are $4.65$ and $4.81$, respectively, and thus smaller than the one for the 
linear instability of the laminar profile.

The flow field of $TW_{TS}$ at the the minimal Reynolds number $Re_{B}=2609$ is visualized in 
figure \ref{fig_BifTSwaves}b).
 
\begin{figure}[t]\vspace*{4pt}
\centering
\includegraphics[]{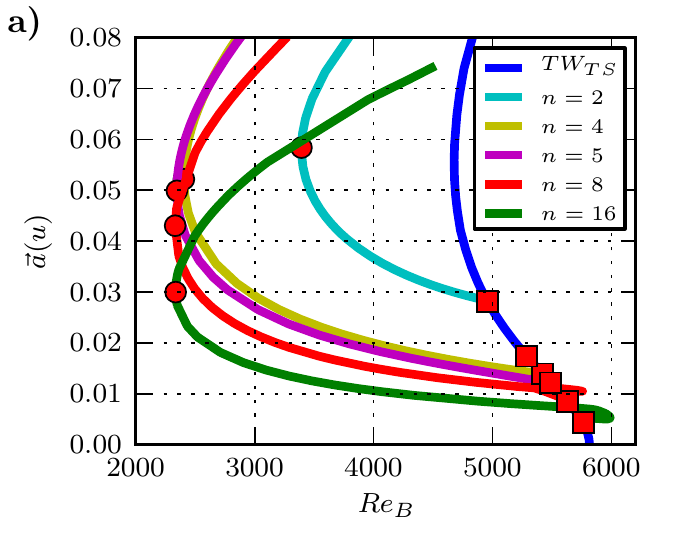}\includegraphics[]{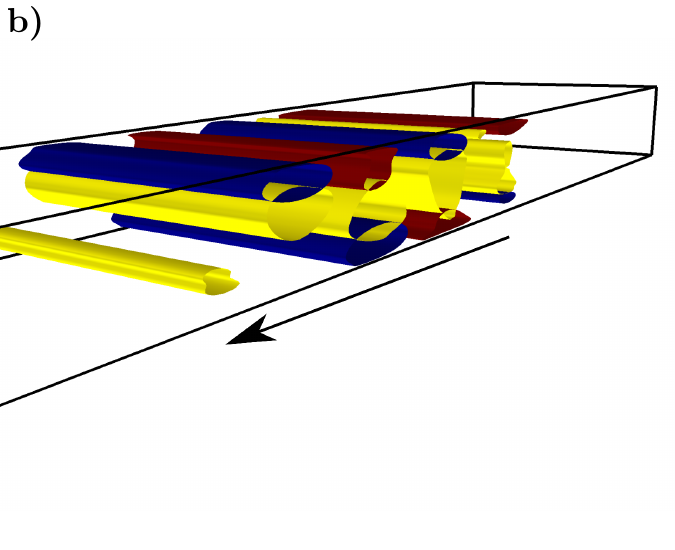}
\caption{Localized TS-waves. (a) Bifurcations off the extended state for
different domain widths. The black line shows the spatial extended TS-wave with streamwise wavelength $\lambda_{0}=2\pi$
and the colored lines
show bifurcating periodic orbits corresponding to streamwise wavelengths $n\lambda_{0}$, where $n$  is integer $2$ and $16$.
The squares indicate the bifurcations of the extended state, and the circles the minimal critical Reynolds numbers for the
different domains, from which a global minimum of $Re_B=2335$ is estimated.
(b) Localized TS-wave with $\L_{x}=16\lambda_{0}$ at the minimum $Re_{B}=2335$.  Isosurfaces of $Q=0.01$ are shown in yellow,
isosurfaces of $u=0.1$ and $u=-0.1$ in red and blue.}\vspace*{-6pt}
\label{fig_BiflocTSwaves}
\end{figure}

In a study on two-dimensional PPF, Jimenez \cite{Jimenez1990a} discovered streamwise modulated packages of TS-waves.
They have a turning point significantly lower than the spatially extended traveling wave $TW_{TS}$ \cite{Price1993}. 
In even longer domains, they turn into fully localized states that are relative periodic orbits, generated out of subharmonic instabilities of the spatially extended TS waves \cite{Drissi1999,Mellibovsky2015}.
Figure \ref{fig_BiflocTSwaves}a) shows a bifurcation diagram for the modulated TS-waves and the spatially extended traveling wave 
from which they bifurcate. The ordinate of the diagram is the amplitude of the flow field,
\begin{equation}
a(\textbf{u}) = \sqrt{\frac{1}{2L_{x}L_{z}}\int \textbf{u}^{2} dx dy dz}, 
\end{equation}
where $L_{x}$ and $L_{z}$ are the streamwise and spanwise wavelengths of the computational domain.
The spatially extended TS-wave shown in the figure has a wavelength $\lambda_{0}=2\pi$ and the  localized and modulated TS-waves are created in Hopf-bifurcations that correspond to streamwise wavelengths $L_{x}$ between 
$4$ to $16$ times $\lambda_{0}$. Just above their bifurcations all solutions are modulated TS-waves but if the wavelength
of the modulation is sufficiently long they become increasingly more localized with decreasing $Re_{B}$.
E.g. for the short modulation wavelength $\Lambda=6 \lambda_{0}$, the solution is modulated but extends across the domain,
whereas for $L_{x}=16 \lambda_{0}$ it becomes well localized at low Reynolds numbers.

The lowest Reynolds numbers for the turning point are $Re_{B}=2335$ and $Re_{P}=2372$ and 
are achieved for the localized TS-wave with modulation wavelength $16\lambda_{0}$. 
Further increasing the modulation wavelength does not change the localized state and does not result in significant changes 
of the Reynolds number for the turning point.

The spatially extended states as well as their localized counter-parts are extremely unstable to three-dimensional disturbances.
However, the lower branches are stable against super-harmonic two-dimensional disturbances. 
The superharmonic bifurcations of the spatially extended TS-waves were investigated in detail by Casas and Jorba \cite{Casas2012}, who found that the upper branch of $TW_{TS}$ undergoes Hopf bifurcations creating periodic state for
the case of constant pressure gradient as well as for constant mass flux.

\begin{figure}
\centering
\includegraphics[]{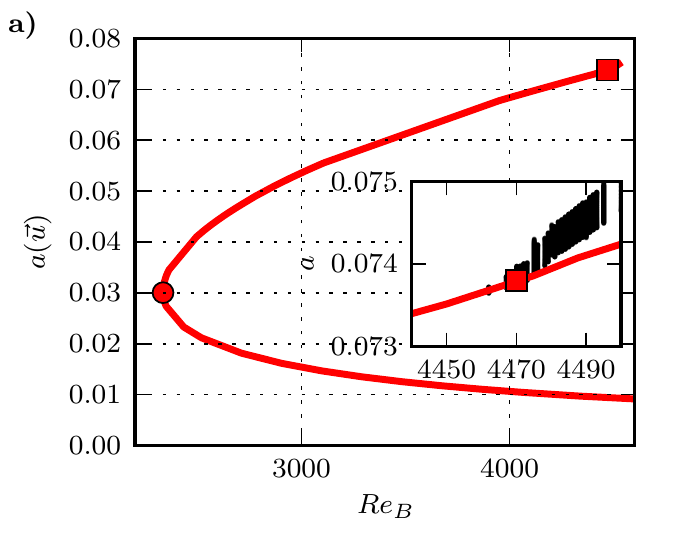}\includegraphics[]{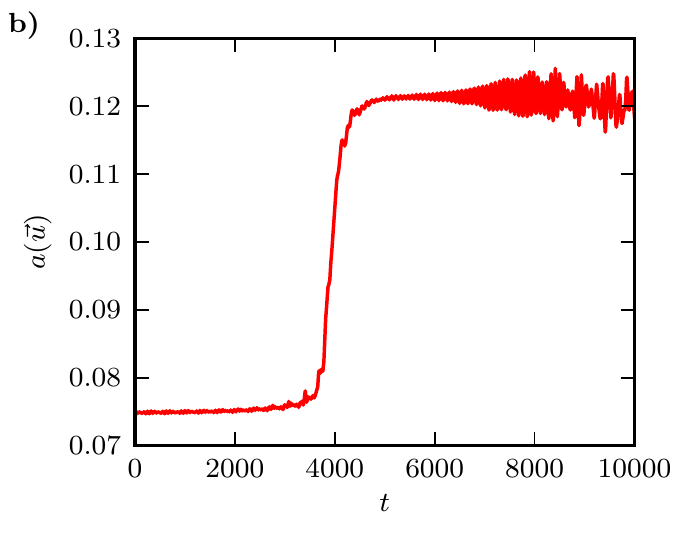}
\caption{Bifurcations of localized TS-waves. (a) The upper branch of the localized TS-wave undergoes a 
Hopf bifurcation at $Rd_B=4470$ (red square) to a state with two frequencies. The states are indicated by the minima in $a$,
which for a periodic orbit gives one point, and for the quasiperiodic one a line (see inset). 
The turning point is marked by a red dot, the bifurcation point by a red disk.
(b) For even higher Reynolds numbers, the localized state undergoes a transition to a spatially extended state, here illustrated
with the time trace of $a(t)$ for a state with $Re_B=4496$. The sharp increase at times $O(4000)$ marks the transition from 
the localized to an extended stated. 
\label{fig_BifDiag_LocTS32pi}}
\end{figure}

The lower branch of the streamwise localized TS-wave has only one unstable eigenvalue in the two-dimensional subspace and thus it is the edge state of the system \cite{Skufca2006,Kreilos2012}.
The upper branches undergo further bifurcations, leading to a chaotic temporal evolution. For the localized TS-wave, 
the upper branch undergoes a Hopf bifurcation and adds another frequency at $Re_{B}=4470$.
Further bifurcations of this quasi-periodic state add additional frequencies leading to a chaotic state.
In figure \ref{fig_BifDiag_LocTS32pi}a) a bifurcation diagram that includes the localized TS-wave and the bifurcating states is given.
The Hopf bifurcation of the upper branch is marked in the figure by a red square.

For $Re_{B}>4493$ the chaotic state becomes unstable and spontaneous transitions to a spatially extended chaotic state are possible. 
An example of a trajectory which spends several thousand time units in the vicinity of
the localized chaotic state before it switches to the spatially extended state is shown in figure \ref{fig_BifDiag_LocTS32pi}b).

\section{Three-dimensional travelling waves}
The first three-dimensional solutions for PPF were described by Ehrenstein \& Koch \cite{Ehrensteint1991}.
They found three-dimensional traveling waves below the saddle-node point of the two-dimensional TS-waves and studied
their dependence on the streamwise and spanwise wavelength. However, their calculation were strongly truncated and attempts to
reproduce their results with better resolution failed \cite{Nagata2013a}. Subsequent studies identified 
many other three-dimensional exact solutions for PPF \cite{Waleffe2001,Nagata2013a,Gibson2014,Zammert2014a,Zammert2014b,Rawat2014}.
Waleffe \cite{Waleffe2003} and Nagata \& Deguchi \cite{Nagata2013a} studied
the dependence of particular solutions on the streamwise and spanwise wavelengths. They give as estimates for the lowest
critical Reynolds numbers for the appearance of a traveling wave the values  $Re_{B}=615$ for $L_{z}=1.55\pi$ and $L_{x}=1.14\pi$
in the case of constant mass flux,
and  $Re_{P}=805.5$ for  $L_{x}=1.504\pi$, $L_{z}=1.156\pi$ in the case of constant pressure drop.
 
\begin{figure}[t]\vspace*{4pt}
\centering
\includegraphics[]{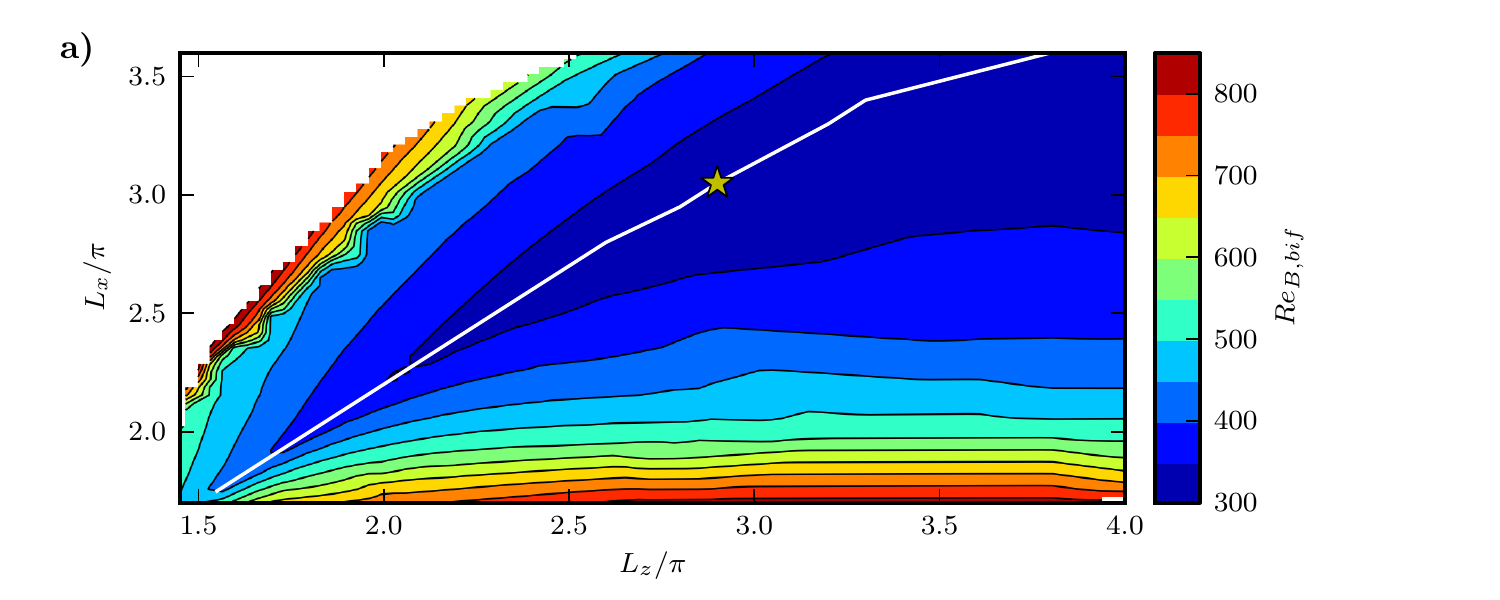}
\\
\includegraphics[]{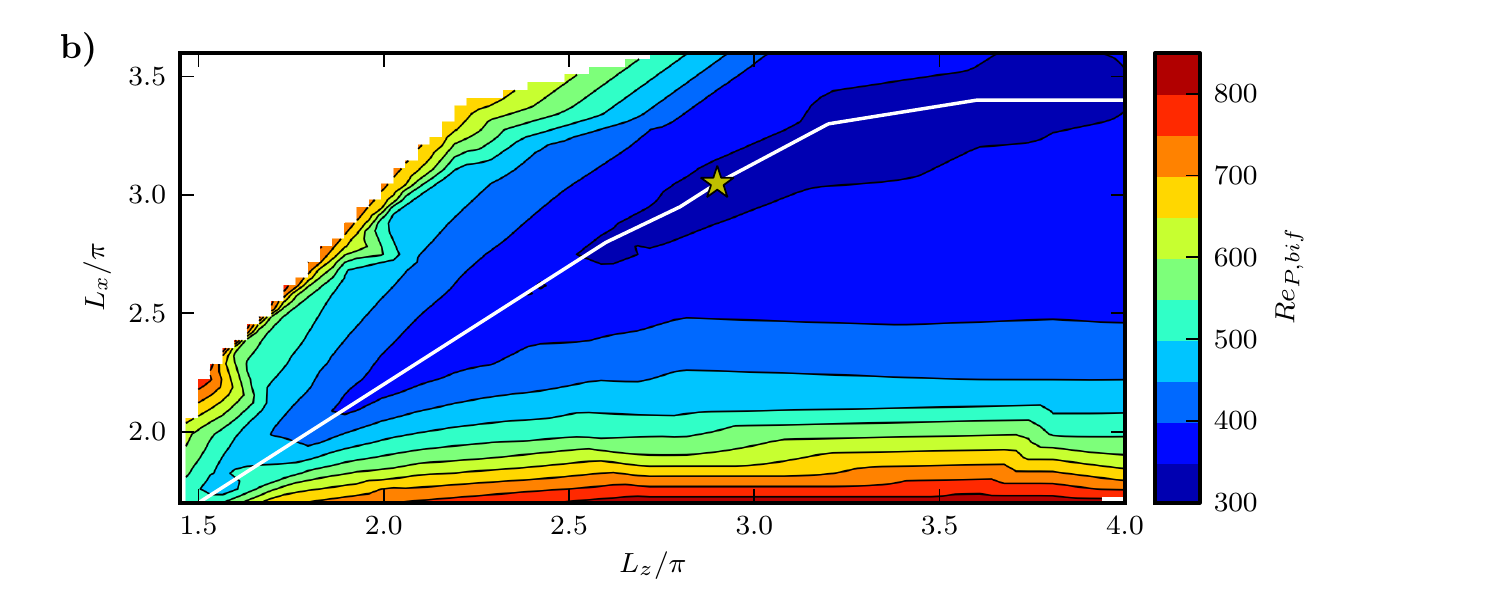}

\caption{Wavelengths dependence of the critical Reynolds numbers for the traveling wave $TW_E$ for (a) constant mean flow and (b) constant pressure drop.
The Reynolds number is color coded and the optimal values are marked by a star. They are
$Re_B=315.8$ for constant mean flow, and $Re_{P}=339.1$ at for constant pressure drop. In both cases the corresponding wavelength are
$L_{x}=2.9\pi$ and $L_{z}=3.05\pi$. The white line marks the optimal $L_x$ for a given
$L_z$.
}\vspace*{-6pt}
\label{fig_ReMinTW}
\end{figure}

In this section, we investigate a special three-dimensional coherent structure, obtained  by tracking the edge state
 \cite{Skufca2006,Schneider2007} in appropriately chosen computational domains. 
We use a numerical resolution of  
$N_{x}\times N_{y} \times N_{z}=32 \times 65 \times 64$ for small domains and increase it 
to $N_{x}\times N_{y} \times N_{z}=80 \times 65 \times 112$ for a domain with $L_{x}=L_{z}=4\pi$.
For sufficiently large domains the edge state of PPF is a traveling wave \cite{Zammert2014b} that is symmetric with respect to the center-plane,
\begin{equation}
 s_{y}: [u,v,w](x,y,z)=[u,-v,w](x,-y,z),
\end{equation}
and obeys a shift-and reflect symmetry in addition,
\begin{equation}
 s_{z}\tau_{x}(\lambda_{x}/2): [u,v,w](x,y,z)=[u,v,-w](x+\lambda_{x}/2,y,-z).
\end{equation}
To obtain the optimal wavelengths for this travelling wave, referred to as $TW_{E}$ in the following, we 
studied the dependence of the critical Reynolds number on the streamwise and spanwise wavelength. 
In a straightforward scan of the wavelength domain with a stepping width of $0.1\pi$ and $0.05\pi$
in the spanwise and streamwise direction, respectively, we obtained the results shown in figure \ref{fig_ReMinTW}.
The optimal Reynolds number is marked by a star. 

\begin{figure}[t]\vspace*{4pt}
\centering
\includegraphics[]{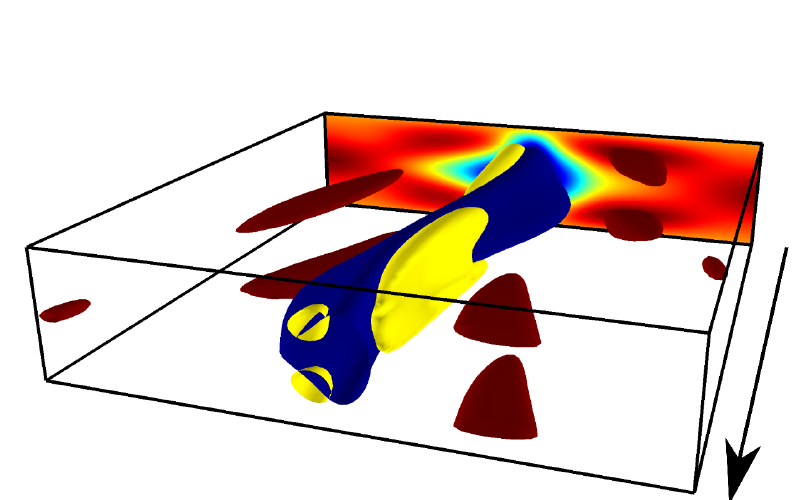}
\caption{Instantaneous snapshot of
$TW_{E}$ for $L_{x}=2.9\pi$  and $L_{z}=3.05\pi$ at the turning point at $Re_{B}=315.8$. Iso-surfaces of $Q=0.01$ are shown in yellow, 
isosurfaces of $u=0.1$, $-0.25$ of the
streamwise velocity component (deviation from the laminar profile) in red and blue.
The direction of the flow is indicated by the black arrow.}\vspace*{-6pt}
\label{fig_YZplaneTW}
\end{figure}

As indicated by the white lines in figure \ref{fig_ReMinTW} , the lowest Reynolds numbers for the 
turning points are achieved if spanwise and streamwise wavelengths are of comparable size
for small wavelengths where the states are not localized in spanwise direction.
In wide domains, for sufficiently large $L_z$, the solution becomes localized in the spanwise direction, 
and the turning point depends on $L_{x}$ only.
The lowest bulk Reynolds number for the turning point is $315.8$ and achieved of $L_{x}=2.9\pi$ and
a width $L_{z}= 3.05\pi$. A visualization of the flow structures for these critical parameter values is given in figure \ref{fig_YZplaneTW}.
The lowest values of the pressure Reynolds number is $Re_{P}=339.1$, which is also achieved for $L_{x}=2.9\pi$ and $L_{z}=3.05\pi$.
Both minimal Reynolds numbers are much lower than those reported in previous studies \cite{Waleffe2003,Nagata2013a}.

The optimal state $TW_E$ may be compared to the coherent states in plane Couette flow. There, the lowest critical 
Reynolds number for the appearance of exact coherent structures is $127.705$ \cite{Clever1997,Waleffe2003}, 
based on half the channel height, and half the velocity difference between the plates.  
PPF is like two plane Couette flows staggered on top of each other, though with different boundary conditions
(free slip instead of rigid) at the interface. For the Reynolds numbers one then has to take into account that the 
mean velocity in PPF corresponds to the full velocity difference between the plates in plane Couette flow, 
and half the channel width corresponds to the full gap. Therefore, when comparing Reynolds numbers, the 
ones from the usual definition of plane Couette flow have to be multiplied by four. Using this redefinition of the 
Reynolds number, the minimal Reynolds number of $Re=127.705$ for plane Couette flow 
corresponds to $Re_{B}=510$ for PPF. The proper steps for this comparison were done taken by
Waleffe \cite{Waleffe2003}, who implemented a homotopy between the two flows, including the boundary conditions.
He finds an optimal Reynolds number $Re_{B}\approx 642$. Moreover, the optimal wavelength for his state are
$L_x=1.86\pi $ and $L_z= 0.74$, much shorter and narrower than for the state described here (see figure \ref{fig_ReMinTW}).
Therefore, for the time being, the travelling wave $TW_E$ described here is the lowest lying critical state for plane
Poiseuille flow.

\begin{figure}
\centering
\includegraphics[]{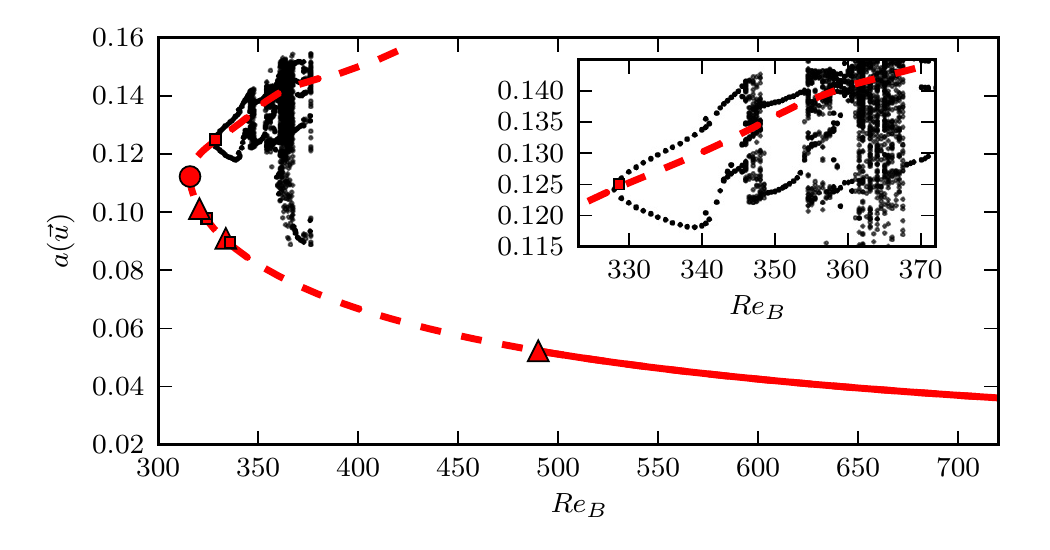}
\caption{Bifurcation diagram of $TW_{E}$ for the spatial wavelengths leading to the lowest bulk Reynolds number for the bifurcation point.
For Reynolds numbers where the wave has only one unstable direction ($Re_{B}>490$) a solid red line
is used, while if the wave has more unstable directions a dashed line is used. 
The turning point is marked by a red dot. Red squares mark the positions of a Hopf bifurcations and triangles
those of pitchfork bifurcations. The black dots
are obtained from maxima and minima during the time-evolution. They indicate further bifurcations and the appearance of temporal chaos.
\label{fig_BifDiag_TWE}}
\end{figure}

For the combination of wavelengths giving the lowest values of $Re_{B}$ for the turning point,
the bifurcation diagram of $TW_{E}$ is shown in figure \ref{fig_BifDiag_TWE}. 
In a subspace with the symmetries $s_{y}$ and $s_{z}\tau_{x}(\lambda_{x}/2)$ the 
the upper branch of $TW_{E}$ is stable for $Re_{B}<328.6$. 
At this Reynolds number the traveling wave undergoes a Hopf bifurcation, creating a periodic orbit which is stable in
the symmetric subspace. In further bifurcation a chaotic attractor is created which remains stable in the subspace.
The attractor is visualized in figure \ref{fig_BifDiag_TWE} by plotting minima and maxima of $a(\vec{u})$ for a trajectory on the attractor. 

\section{Streamwise localized three-dimensional periodic orbits}
\begin{figure}
\centering
\includegraphics[]{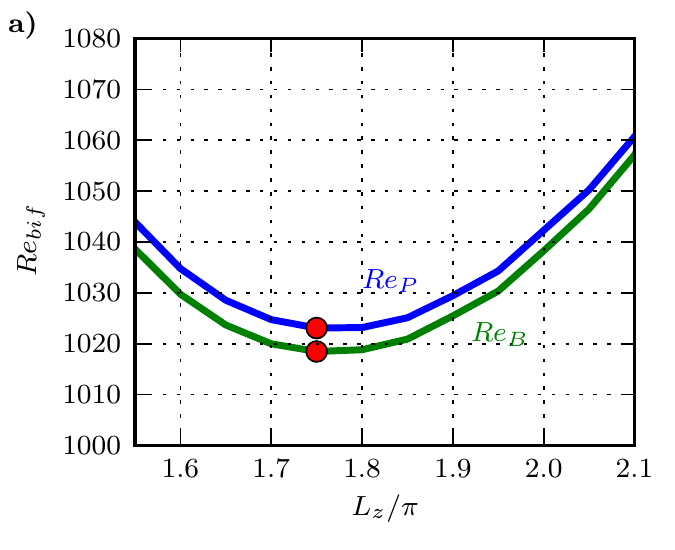}\includegraphics[]{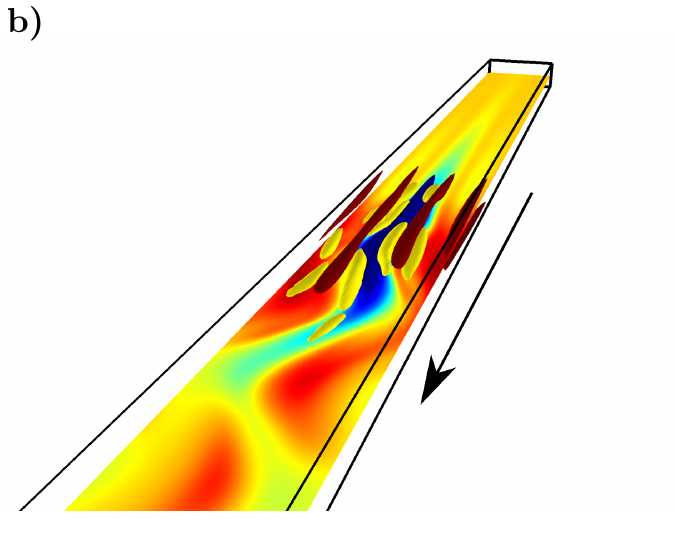}

\caption{The localized periodic orbit $PO_{E}$.
(a) Reynolds number of  the bifurcation point of $PO_{E}$ in dependence on the spanwise wavenumber $L_{z}$.
Note that while the critical Reynolds numbers again vary between constant mean flow and constant pressure drop,
but the optimal wavelengths coincide.
(b) Instantaneous snapshot of $PO_{E}$ for $L_{z}=1.75\pi$ and 
$Re_{B}=Re_{B,min}(PO_{E})=1018.5$. Isosurfaces of $Q=0.002$ are shown in yellow and isosurfaces $u=0.075$, $-0.15$ of the
streamwise velocity component (deviation from the laminar profile) in red and blue. The direction of the flow is indicated by the black arrow.
In the center-plane the streamwise velocity is color-coded from blue to red. }
\label{fig_POE}    
\end{figure}

In spatially extended computational domains the edge state $TW_E$ undergoes long-wavelength instabilities
creating modulated and localized exact solutions \cite{Melnikov2013}. In particular,
a subcritical long-wavelength instability of $TW_{E}$ at high Reynolds numbers creates a streamwise localized periodic orbit, henceforth referred to as $PO_{E}$,
which is an edge state in long computational domains \cite{Zammert2014b} and appears at  low Reynolds numbers in a saddle-node bifurcation.
The dependence of the Reynolds number of this saddle-node bifurcation on the spanwise wavelength is shown in figure
\ref{fig_POE}a).The lowest Reynolds number for the turning point
is achieved for a spanwise wavelength of $1.75\pi$. For this value of $L_{z}$ the bifurcation point
lies at $Re_{B}=1018.5$ and $Re_{P}=1023.18$, respectively. 
For large values of $L_{z}$ also the orbit 
$PO_{E}$ becomes localized in spanwise direction \cite{Zammert2014b}, 
but for this spanwise wavelengths the bifurcation point lies at a much higher Reynolds number. 
A visualization of $PO_{E}$ for the minimal bulk Reynolds number is shown in  figure \ref{fig_POE}b).

While the localized TS waves, which also arise out of a subharmonic instability of a spatially extended state, 
exist at lower Reynolds numbers than the corresponding  spatially extended solutions,
the localized orbit $PO_{E}$ appears at much higher Reynolds number than its corresponding spatially extended solution $TW_{E}$. 

\section{Summary and conclusion}
The different states and their critical Reynolds numbers and wavelengths are summarized in table 
\ref{tab1}. It is interesting to see that for the TS-waves, the localized structures appear before
the extended ones, whereas for the 3d states the localized ones have a higher critical Reynolds number.
Moreover, all TS waves appear well above the experimentally observed onset of turbulence, near $Re_B\approx 1000$  \cite{Henningson}.

\begin{table}
\tbl{Critical values for the different bifurcations in PPF. The columns give the 
critical Reynolds numbers $Re_{B}$ and $Re_{P}$ and the associated optimal wavelengths
for exact coherent structures and the laminar profile.}
{
 \begin{tabular}{cccc|cccc}
\toprule
               & $Re_{B,min}$ & $L_{x}$    & $L_{z}$ & $Re_{P,min}$& $Re_{\tau,min}$ & $L_{x}$ & $L_{z}$  \\
\colrule
$TW_{TS}$      &   $2610$       &  $1.48\pi$ & -       &2941&   76.7        &$1.53\pi$&-         \\
Localized TS  & 2334         &  -         &      -  &2373 &  68.9       &-        &-        \\
$TW_{E}$       &  315.8       &  $3.05\pi$ &$2.9\pi$ &339.1   &26.04      &$3.05\pi$& $2.9\pi$\\
$PO_{E}$       &  1018.5      &  -         &$1.75\pi$&  1023.18  &45.23   &  -      & $1.75\pi$\\
\colrule
Laminar state &  5772.22 & $1.96\pi$&-         &5722.22&196.98&$1.96\pi$&   -    \\
\botrule
\end{tabular}
}
\label{tab1}
\end{table}

The traveling wave $TW_{E}$ appears at very low Reynolds number, well below the experimentally 
observed onset of turbulence. 
The bifurcation diagram in figure \ref{fig_BifDiag_TWE} 
shows the usual increase in complexity and the presence of a crisis bifurcation, in which the attractor turns 
into a repellor and the dynamics becomes transient. 
In addition to the states discussed here, there are a variety of bifurcations in which  
spanwise localized and doubly-localized solution branch off from $TW_E$. 
All of these states, as well as exact coherent structures with different flow fields contribute to the
temporal evolution, and the network that forms with increasing Reynolds numbers and that can then
carry the turbulent dynamics.

Very little is known about the lowest possible Reynolds number for the appearance of coherent structures,
and there are hardly any methods for determining them reliably \cite{Pausch2014a,Chernyshenko:2014je,Huang:2015jp}
However, the state $TW_E$ has the potential to be the lowest possible state in PPF flow by analogy to the lowest
lying state in plane Couette flow. Identifcation of lower lying exact coherent structures in either flow should therefore 
also have implications for the other flow.


Among the exact coherent structures,  $TW_{E}$ is a harbinger for the occurrence of turbulence, whereas
the two dimensional TS-wave are latecomers. 
They do not contribute to the formation of subcritical transition at low Reynolds, 
but are related to a secondary path to turbulence in PPF that can be realized if special precautions 
are taken to prevent the faster transition via 3d structures. 

\vspace{0.5cm}

This work was supported in part by the DFG within FOR 1182.

\bibliographystyle{tJOT}

\end{document}